\documentclass[fleqn,usenatbib]{mnras}

\usepackage{newtxtext,newtxmath}

\usepackage[T1]{fontenc}
\usepackage{ulem}
\usepackage{xcolor}
\DeclareRobustCommand{\VAN}[3]{#2}
\let\VANthebibliography\thebibliography
\def\thebibliography{\DeclareRobustCommand{\VAN}[3]{##3}\VANthebibliography}

\usepackage{graphicx}	
\usepackage{amsmath}	
\usepackage{physics}

\title[Radio emission in Abell 85]{On the Origin of Diffuse Radio Emission in Abell 85 - Insights from new GMRT Observations}
\author[Rahaman et al.]{
Majidul Rahaman$^{1}$\thanks{E-mail: phd1601121007@iiti.ac.in, rmajidul@gmail.com},
Ramij Raja$^{1}$,
Abhirup Datta$^{1}$,
Jack O Burns$^{2}$,
and David Rapetti$^{3,4,2}$ \\
$^{1}$Department of Astronomy, Astrophysics and Space Engineering, \href{http://www.iiti.ac.in/}{Indian Institute Of Technology Indore}, Indore, India\\
$^{2}$Center for Astrophysics and Space Astronomy, Department of Astrophysical \& Planterary Sciences and Department of Physics, University of Colorado Boulder, \\ Boulder, CO 80309, USA\\
$^{3}$NASA Ames Research Center, Moffett Field, CA 94035, USA\\
$^{4}$Research Institute for Advanced Computer Science, Universities Space Research Association, Columbia, MD 21046, USA\\}

\date{Accepted 2022 June 29. Received 2022 June 29; in original form 2021 March 29}

\pubyear{2022}

\begin{document}
\label{firstpage}
\pagerange{\pageref{firstpage}--\pageref{lastpage}}
\maketitle
\begin{abstract}
Extended, steep, and ultra-steep spectrum radio emission in a galaxy cluster is usually associated with recent mergers.
Simulations show that radio phoenixes are aged radio galaxy lobes whose emission reactivates when a low Mach shock compresses it.
A85 hosts a textbook example of a radio phoenix at about 320 kpc southwest of the cluster center.
We present a new high resolution 325 MHz GMRT radio map illustrating this radio phoenix's complex and filamentary structure.
The full extent of the radio structure is revealed for the first time from these radio images of A85.
Using archival \textit{Chandra} X-ray observations,
we applied an automated 2-D shock finder to the X-ray surface brightness and Adaptive Circular Binning (ACB) temperature maps which confirmed a bow shock at the location of the radio phoenix.
We also compared the Mach number from the X-ray data with the radio-derived Mach number in the same region using multi-frequency radio observations and find that they are consistent within the 1$\sigma$ error level. 
\end{abstract}


\section{Introduction} \label{sec:Intro}
Galaxy clusters are the largest gravitationally bound, organized objects in the Universe. Intersecting cosmic filaments of several megaparsecs collapse and form clusters \citep{Rosati2002}. 
Clusters of galaxies contain the most of its baryons in the intracluster medium (ICM), which is hot dilute gas ($\sim10^{-3}\mathrm{cm}^{-3}$) at $10^7-10^8$ K temperature and have a magnetic fields of 0.1-10 $\mu$G \citep{van_Weeren2019SSRv..215...16V}. 
This magnetized hot plasma emits X-ray radiation by thermal Bremsstrahlung throughout the cluster volume. 
Based on radio observations, it is clear that the ICM can also contain a non-thermal component of cosmic rays (CR), which are not directly connected to the radio galaxies present within the cluster. 
Cluster mergers create shocks and turbulence in the ICM, which amplify the ICM magnetic fields and re-accelerate relativistic electrons. 
In the presence of $\mu$G ICM magnetic fields, these high energy relativistic particles emit large scale synchrotron radiation in the form of relics, radio halos, and mini-halos \citep{Brunetti2014IJMPD..2330007B,van_Weeren2019SSRv..215...16V}. 
The cluster-wide diffuse radio sources called radio haloes are believed to originate from the re-acceleration of \textit{in-situ} electrons via turbulence created during cluster mergers \citep{Brunetti2014IJMPD..2330007B}. 
Relics are diffuse extended (size $\sim1$ Mpc) or roundish radio sources, with a steep spectrum, and strongly polarized (10-30\%), located at the cluster periphery and believed to be associated with merger shocks \citep{Feretti2012}.  \citet{Kempner2004rcfg.proc..335K} distinguished radio relics in two classes: those related to previous AGN activity called a radio Phoenix and those related to the ICM called a radio Gischt.
Radio Gischt are associated with cluster radio shocks \citep{van_Weeren2019SSRv..215...16V}.

Radio phoenixes are a less studied class of diffuse radio sources in the cluster environment. 
The currently favored scenario is that phoenixes are a manifestation of fossil plasma in galaxy clusters from past episodes of AGN activity \citep{van_Weeren2019SSRv..215...16V}.
Simulations by \citet{Enblin2001A&A...366...26E} show that when merger shocks compress this fossil plasma, the momentum of the relativistic electrons and magnetic field strength increase, which results in synchrotron radiation characterized by an ultra-steep and curved radio spectrum. 
By this mechanism, the relativistic electrons can be maintained at higher energies than what radiative cooling alone would allow. \citet{Enblin2002MNRAS.331.1011E} also predict that sources originating from this compression scenario should have complex and filamentary morphology. 
The formation scenario for these revived fossil plasma sources is still unclear because of the lack of observational evidence supporting those scenarios \citep{van_Weeren2019SSRv..215...16V,Mandal_a2019A&A...622A..22M}.

In a recent study,  \citet{ZuHone_2021ApJ...914...73Z} explored the possibility that radio relics trace the shape of the underlying distribution of seed relativistic electrons that lit up by a recent shock passage. \citet{ZuHone_2021ApJ...914...73Z} used magneto-hydrodynamic (MHD) simulations of cluster mergers and included bubbles of relativistic electrons introduced by a central AGN jet or a radio galaxy from an off-center region.
They found that merger-driven gas motions can advect and spread relativistic seed electrons creating extended, filamentary, or sheet-like intracluster plasma regions enriched with aged cosmic rays matching radio relics/radio phoenixes. 

A85 is an X-ray luminous cool core cluster (z = 0.0556, \citealt{Pislar1997,Oegerle2001}). 
A85 hosts a small scale radio relic, which was classified as a radio phoenix by \citet{Kempner2004rcfg.proc..335K}. 
A high fidelity 325 MHz GMRT radio image presented in this study reveals the complex and filamentary structure of the radio phoenix (see Figure \ref{fig:sw_circle}). 
The thermodynamic structure of the cluster was discussed in the previous study by \cite{Schenck2014} and by \cite{Ichinohe2015} using both \textit{Chandra} and XMM-Newton observations. \citet{Ichinohe2015} also reported a shock front near the SW-subcluster.
Here, in this study, we used \textit{Chandra} X-ray observations 
to produce temperature maps using a different technique disclosing more detailed thermodynamic structure throughout the cluster.
We construct a high resolution temperature map (Figure \ref{fig:T_acb}) using ACB method (\citealt{Sanders2001,Datta2014}) and apply an automated shock finder \citep{Datta2014} with which we confirm the bow shock near to the radio phoenix location (bottom left panel of Figure \ref{fig:sw_circle}). A85 also hosts a large scale gas sloshing arm ($\sim$ 600 kpc, \citealt{Ichinohe2015}), which affects the ICM distribution all over the cluster.
Hence, A85 allows us to investigate the possible connection between the radio phoenix and the X-ray bow shock, gas sloshing as suggested by previous simulations described above (e.g., \citealt{Enblin2001A&A...366...26E,Enblin2002MNRAS.331.1011E,ZuHone_2021ApJ...914...73Z}).

This paper is organized as follows. In Sect. \ref{data_analysis}, we discuss the data analysis of \textit{Chandra} X-ray, 325 MHz GMRT radio and 1.4 GHz VLA radio observations.
In Sect. \ref{results}, we describe the results obtained from both X-ray and radio observations.
In Sect. \ref{discussion}, we discuss the bow shock and its connection to the origin of the radio phoenix. Finally, the conclusions are presented in Sect. \ref{sec:conclude}.

Throughout the paper, we use a $\Lambda$CDM cosmology with $H_0 = 70.0$ km s$^{-1}$ Mpc$^{-1}$, $\Omega_{m} = 0.3$ and $\Omega_\Lambda = 0.7$. Errors are quoted at the 1$\sigma$ level. At the cluster redshift $z = 0.055$, $1\arcsec$ corresponds to a physical scale of 1.12 kpc.

\begin{table}
	\centering
	\caption{Archival \textit{Chandra} X-ray observations (ACIS-I) summary.}
	\label{tab:data_table}
	\resizebox{\columnwidth}{!}{
	\begin{tabular}{lcccr}
		\hline
 		ObsId & Observation & RA & Dec & Clean \\
        & Dates &  &  & Exp (ks) \\
		\hline
		904 & 2000-08-19 & 00:41:45.80 & -09:22:45.00  & 36\\
		15173 & 2013-08-14 & 00:41:41.96 & -09:20:51.65 & 40\\
		15174 & 2013-08-17 & 00:41:53.26 & -09:24:34.03 & 38\\
		16263 & 2013-08-09 & 00:41:53.26 & -09:24:34.03 & 37\\
		16264 & 2013-08-10 & 00:41:41.96 & -09:20:51.65 & 35\\
        \hline
        Total & & & & 186 \\
        \hline
	\end{tabular}
	}
\end{table}

\section{Observations and data analysis}    \label{data_analysis}

\subsection{Chandra X-ray observations}  \label{sec:X-ray}
We used five separate observations of A85 from the archive of the \textit{Chandra} X-ray observatory. Table \ref{tab:data_table} shows the details of each pointing.
All the observations were taken using the ACIS-I detector. One of the observations (ObsID 904) is in the FAINT mode while the rest are in VFAINT mode.
We used a pipeline described in \citet{Datta2014,Schenck2014,Hallman2018}, written in IDL and bash scripts\footnote{Python version of this pipeline also been released as {\it ClusterPyXT}https://github.com/bcalden/ClusterPyXT \citep{Alden_2019ascl.soft05022A}.}. 
This pipeline is semi-automated and designed such that it takes \textit{Chandra} observation Id's and generates high fidelity temperature maps using ACB, and Weighted Voronoi Tessellation (WVT).
The \textit{Chandra} data were calibrated and reprocessed for all five A85 ACIS-I observations with CIAO-4.9 and CALDB 4.7.9. Backgrounds were taken into account using blank-sky backgrounds from CALDB. The background data were processed and reprojected to match the observational data. CTI (Charge transfer inefficiency) corrections (using \textit{acis\_process\_events}) and background flaring were excluded using light curves in the  9.0-12.0 keV and full energy band. The light curves were binned at 259.3 s per bin, and the same binning was also used for blank-sky backgrounds. Count rates greater than $3\sigma$ from the mean were removed using deflare. The light curves were visually inspected to ensure flares were effectively removed. Point sources were detected first by using the \textit{wavdetect} tool in 0.5-1.2 keV, 1.2-2.0 keV, 2.0-7.0 keV energy bands and then visually inspected for false detection. Once \textit{DS9} regions are given to the pipeline, it removes all point sources. The regions that enclosed point sources were also removed from the backgrounds to prevent over-subtracting the background. The above steps were followed to create CLEAN data and background files for each observation separately.

\subsubsection{X-ray Surface Brightness and Temperature Maps Generation} \label{Tmap}
\begin{figure}
\centering
\includegraphics[width=\columnwidth]{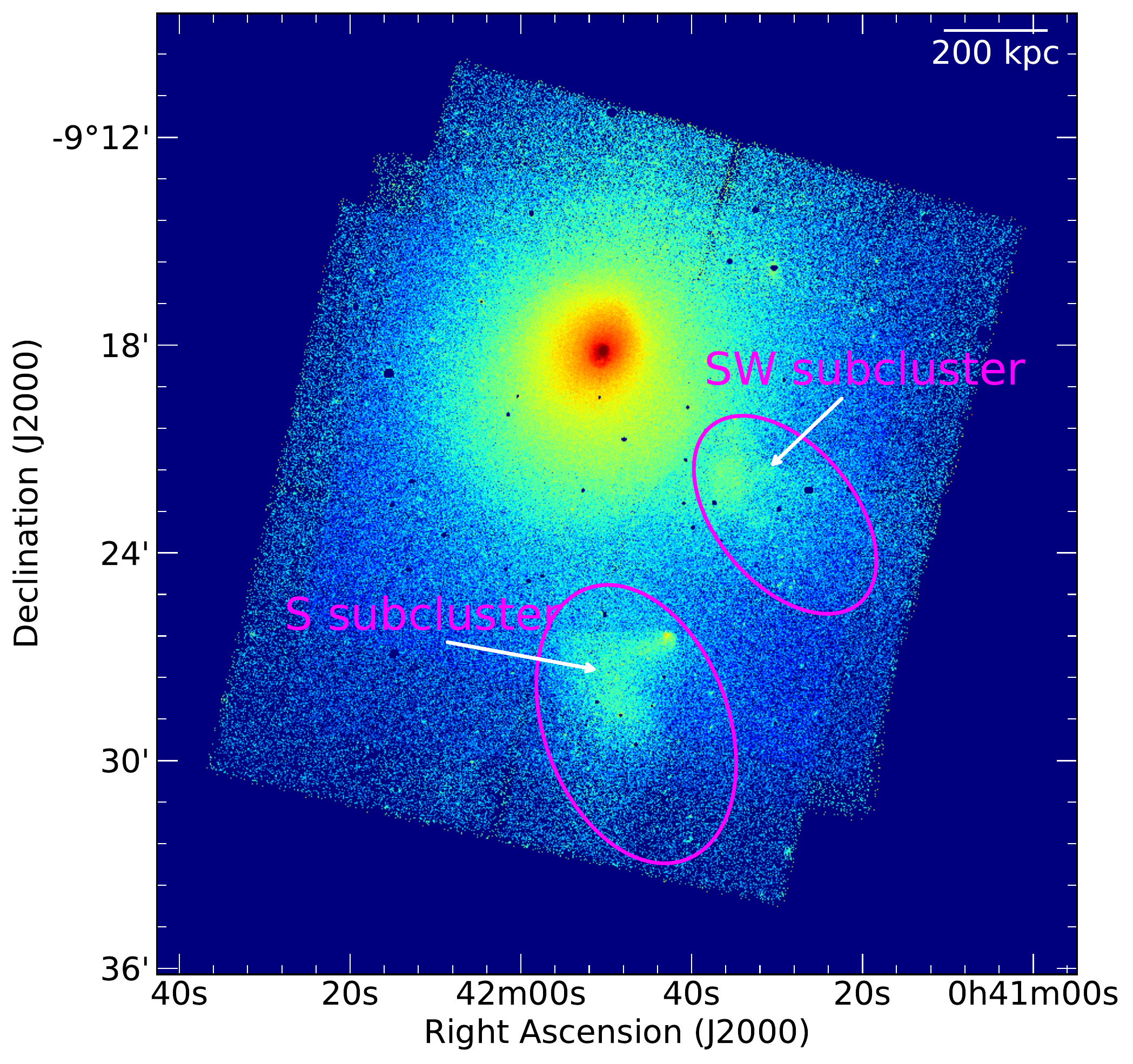}
\caption{Exposure corrected, background subtracted, point source removed surface brightness map of A85 within the 0.7-8.0 keV energy range.
Two elliptical regions (magenta) shows the subcluster regions which were masked during the creation of 2D beta model.}
\label{fig:X_ray_SB}
\end{figure}

After cleaning the data, we combined all data files (5 ObsId; see Table \ref{tab:data_table}) using \textit{merge\_obs} to produce a surface brightness map. We binned CLEAN data by a factor of 4 to produce surface brightness maps. The exposure corrected, background, and point sources removed, 0.7-8.0 keV filtered surface brightness image is shown in Figure \ref{fig:X_ray_SB}.

In order to create the best quality temperature map of a cluster of galaxies, we applied two different techniques: (1) Adaptive Circular Binning (ACB), and (2) Weighted Voronoi Tessellation (WVT). We discuss WVT in Appendix  \ref{wvt}.
The ACB temperature map was produced using the method described in \citet{Randall2008,Randall2010}. The source and background spectra were extracted from the circular regions corresponding to every pixel.
In this process, circles are allowed to overlap with each other.
We used a threshold signal-to-noise ratio (SNR) as a criterion to create circular regions having a minimum of 1000 counts per region or bins for each of the methods. 
Here, the signal is the CLEAN data, and the noise is assumed to be Poisson distributed with contributions from the CLEAN data and blank-sky background.
After creating regions, we used the \textit{dmextract} \textit{CIAO} task to extract spectra from both source and background for each observation separately for the same region. The weighted response matrix (RMF) and weighted effective area (ARF) were calculated using \textit{specextract} for each observation.

Temperature maps were created using the APEC thermal plasma model and photoelectric absorption model PHABS. 
APEC is an emission spectrum from collisionally-ionized diffuse hot gas. We used redshift z = 0.055, metallicity = 0.3$Z_\odot$ \citep{Werner_2013Natur.502..656W,Su_2017ApJ...851...69S}, where $Z_\odot$ is the solar abundance in \citet{Anders_1989GeCoA..53..197A} and $ N_H=2.8 \times 10^{20}\ \mathrm{ cm^{-2}}$ \citep{Kalberla_2005A&A...440..775K}.
We performed spectral fitting using XSPEC version 12.9.1 within the 0.7-8.0 keV energy range. The APEC and PHABS models were fitted to all five spectra for each region from 5 ObsId's simultaneously. C-Statistics \citep{Cash1979} was used for spectral fitting.
The metallicity, redshift, and $N_H$ of the cluster were kept frozen during the fitting process due to the limited number of x-ray counts throughout much of the cluster.
Only APEC normalization and temperature were fitted for each region.
The best-fitted temperature and errors of each circular spectral regions were assigned to the center of the circle. The ACB temperature map and the corresponding error map are presented in Figure \ref{fig:T_acb}.

\begin{table*}
	\centering
	\caption{Summary of archival radio observations of A85.}
	\label{tab:Rdio_data_table}
	\resizebox{\textwidth}{!}{
	\begin{tabular}{lccccr}
		\hline
 		Array & Project &  Frequency & Bandwidth & Observation & Obs. Time \\
        & ID & (MHz) & & Dates & (min)  \\
		\hline
		GMRT & 30\_085 & 325 & 32 MHz & July 28, 2016 & 595 \\
		VLA (B) & AS640 & 1400 & 50 MHz & September 1, 1998 & 540   \\
		\hline
	\end{tabular}
	}
\end{table*}

\subsection{Radio Observations} \label{sec:radio}
To study the nature of the diffuse radio structure of A85, we analyzed unpublished archival GMRT observations at 325 MHz.  We also reanalyzed 1.4 GHz radio observations from the VLA \citep{Schenck2014}.

\subsubsection{1.4 GHz VLA observation}
The VLA data were taken in B configuration on September 1, 2001, and were previously presented by \citet{Slee2001} and \citet{Schenck2014}. The data reduction was carried out using the Astronomical Image Processing System (AIPS) following standard calibration methods. Widefield imaging was performed using 31 facets, and the Briggs robust parameter \citep{Briggs1995} was chosen to be 1. 
Several rounds of phase-only self-calibration were performed to correct for residual phase errors, and finally, the image was restored with a restoring beam of $10\arcsec \times 9\arcsec$. The off-source rms noise achieved is about 15.3 $\mu$Jy beam$^{-1}$, and around the southwest radio emission is about 30 $\mu$Jy beam$^{-1}$.

\subsubsection{325 MHz GMRT observation}
The GMRT 325 MHz observation (PI: Stephen Hamer, July 28, 2016) was performed with 32 MHz bandwidth in dual-polarization mode.
The data reduction and imaging were carried out using SPAM. SPAM is a fully automated python-based pipeline that employs AIPS for reducing data \citep{Intema2009A&A...501.1185I,Intema2017A&A...598A..78I}. With direction-dependent calibration and image plane ripple suppression capabilities, it produces high-resolution radio images at low frequencies. 
Figure \ref{fig:sw_circle} shows the 325 MHz radio image of A85 with off-source rms noise of 70 $\mu$Jy beam$^{-1}$ and near the southwest radio emission of about 150 $\mu$Jy beam$^{-1}$ with a resolution of $10\arcsec \times 9\arcsec$.

The integrated flux density ($S_{\nu}$)\footnote{$S_{\nu} \propto \nu^{\alpha}$, where $S_{\nu}$ is the flux density at frequency $\nu$ and $\alpha$ is\\ the spectral index.} of each radio source was calculated within the 3$\sigma$ contour, where $\sigma$ is the local \textit{rms} background noise.
The error in the total flux density was calculated as $\sqrt{(\sigma_{cal} \times S_{\nu})^2 + (\sigma_{rms} \times \sqrt{N})^2}$, where $\sigma_{cal}$ is the calibration error, $\sigma_{rms}$ is the local \textit{rms} background noise and N is the total number of beams within the 3$\sigma$ contour.
We assumed a calibration uncertainty of 10\% for both GMRT and VLA observations.

The L-band VLA observations have the uv-range of $\approx 0.3 - 54 k\lambda$ and the GMRT observations have the uv-range of $\approx 0.1 - 30 k\lambda$. Therefore, we choose a similar uv-range of $0.5 - 30 k\lambda$ to calculate the spectral index values.

\begin{figure*}
\centering 
\begin{tabular}{lccr} 
    \includegraphics[width=\columnwidth]{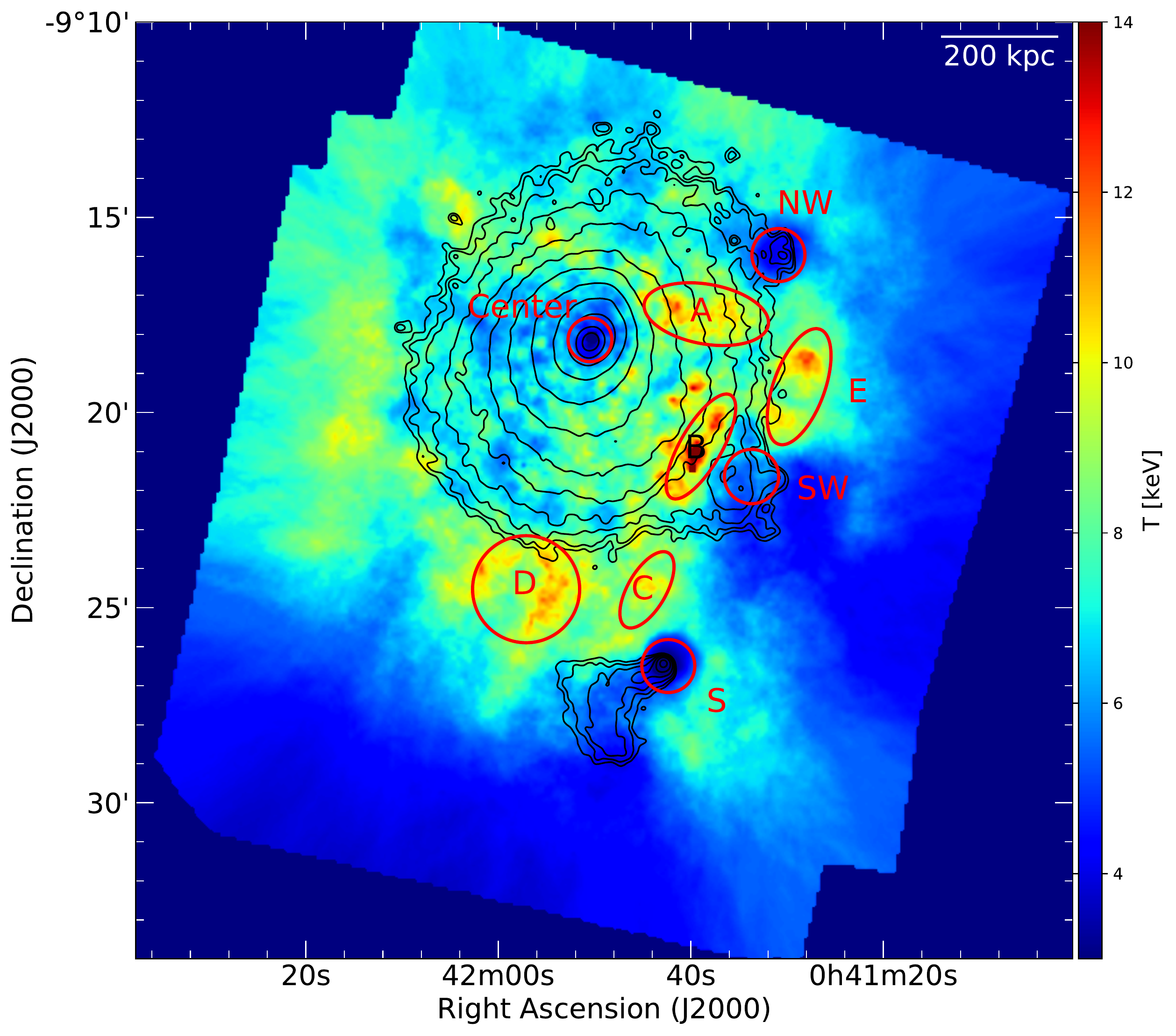} & 
    \includegraphics[width=\columnwidth]{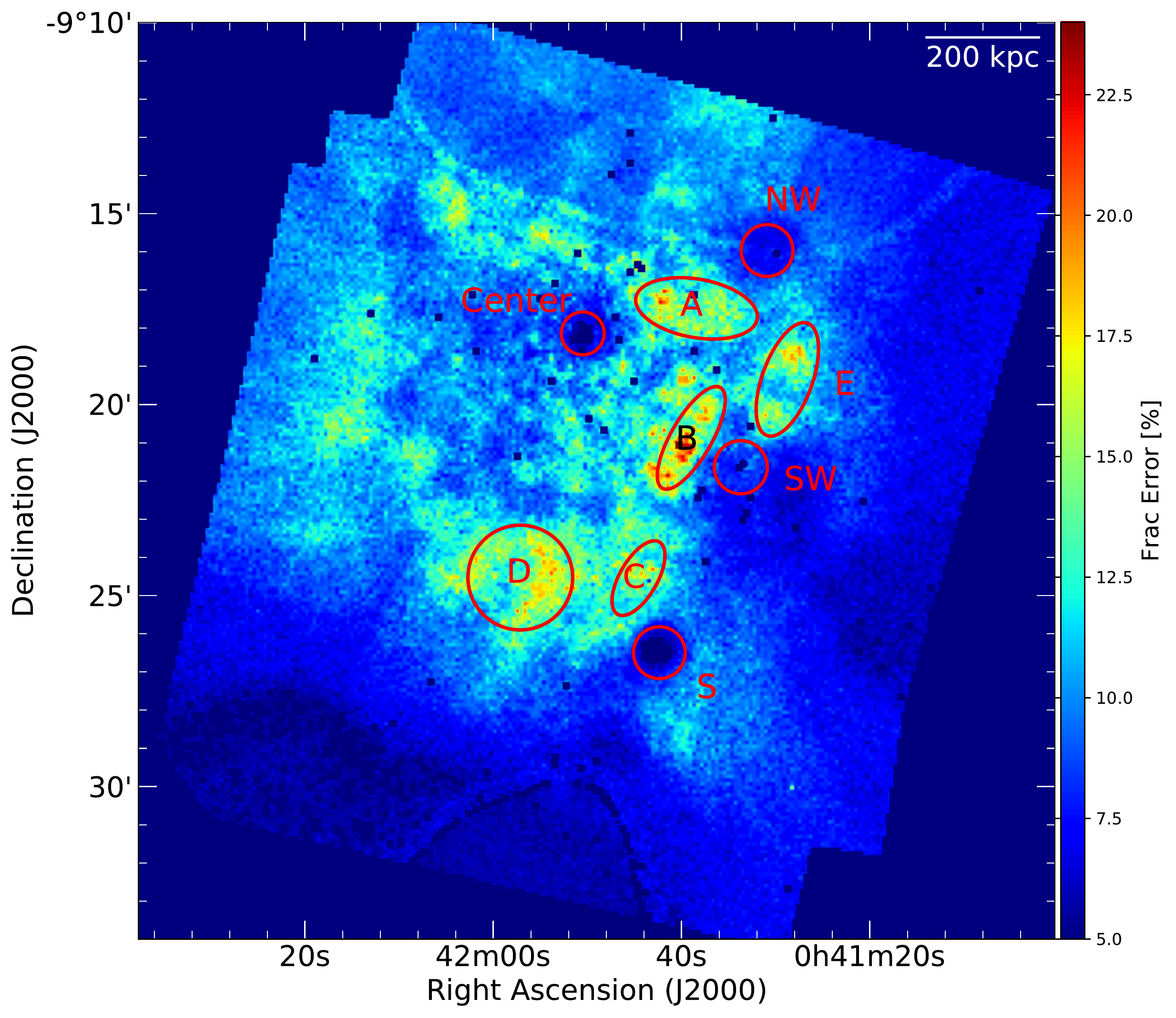} \\ 
\end{tabular} 

\caption{Left: \textit{Chandra} X-ray ACB temperature map of A85 overlaid with \textit{Chandra} X-ray surface brightness (0.7 - 8.0 keV) contours (black). This map reveals a cool core cluster with core temperature $\sim$ 4.7 keV. Average temperatures from different labelled regions (red) are listed in Table \ref{tab:comp_table} which compares them to those obtained from the WVT temperature map.
Right: Fractional error map corresponding to the ACB temperature map within $1\sigma$ level.The regions (A-E) are same as in the left panel.
}
\label{fig:T_acb}
\end{figure*}

\section{Results} \label{results}
\subsection{Temperature maps}
Figure \ref{fig:X_ray_SB} shows the point source removed, background subtracted, projected \textit{Chandra} X-ray surface brightness map of A85, in which A85 appears relaxed, except for the gas clumps in the south and southwest corresponding to two well-studied subclusters \citep{Durret2003,Durret2005,Tanaka2010,Schenck2014,Ichinohe2015}.

In Figure \ref{fig:T_acb}, we display our high fidelity \textit{Chandra} X-ray ACB temperature map of A85.
This map shows a cool core cluster with a disturbed morphology along with several hot as well as comparatively cool regions. 
The region labeled S represents the cool core of the south subcluster, and C and D are regions shock heated by the shock created by the south subcluster \citep{Schenck2014,Ichinohe2015}. 
Regions B \citep{Ichinohe2015} and E \citep{Schenck2014,Ichinohe2015} are also shock heated by the supersonic passage of the southwest subcluster through the ICM. 
The temperature structure of A85 is complex and confirms that the cluster is experiencing more than one massive and one small merger events at the periphery of the cluster.

A85 is a cool core cluster with the core temperature being $4.7 \pm 0.3$ keV, which is about $\sim3-4$ times lower than the surrounding temperature.
As a quantitative comparison between the ACB and WVT maps (the WVT map is discussed in Appendix  \ref{wvt}), the average temperatures from the regions labeled in Figure \ref{fig:T_acb} are listed in Table \ref{tab:comp_table}. All the temperature values are consistent with each other within the error bars except for region S. The temperature value from region S is slightly higher from the WVT temperature map as it shares a small portion of the comparatively hot WVT bin at the northern side of the region.
The overall temperature structure is also consistent with the previous studies by \citet{Schenck2014,Ichinohe2015}.
We used both surface brightness and ACB temperature maps to produce a shock or Mach number map as described in Sect. \ref{shock_finder}.

\begin{table}
	\centering
	\caption{Comparison of temperatures corresponding to the regions in the cluster shown in Figure \ref{fig:T_acb} from two different temperature map-making methods. Errors are shown at the 1$\sigma$ level.}
	\label{tab:comp_table} 
	\resizebox{\columnwidth}{!}{
	\begin{tabular}{lcr}
	\hline
 	Regions & ACB & WVT \\
 		 & (keV) & (keV) \\
		\hline
        Center (Core) & $4.7 \pm 0.3$ & $4.8 \pm 0.4$\\
        S & $4.0 \pm 0.2$ & $5.2 \pm 0.5$ \\
        SW & $5.8 \pm 0.5$ & $5.7 \pm 0.5$ \\
        NW & $4.7 \pm 0.3$ & $5.3 \pm 0.5$ \\
        A & $9.4 \pm 1.2$ & $9.3 \pm 1.4$ \\
        B & $10.3 \pm 1.5$ & $10.4 \pm 1.7$ \\
        C & $9.0 \pm 1.2$ & $10.2 \pm 1.5$ \\
        D & $9.5 \pm 1.3$ & $9.9 \pm 1.6$ \\
        E & $9.4 \pm 1.2$ & $9.0 \pm 1.3$ \\
        \hline
	\end{tabular}
	}
\end{table}

\subsection{Residual map} \label{residual}
After removing the best-fitting 2D elliptical beta model to the large-scale cluster X-ray surface brightness, we created relative deviation images to emphasize the small azimuthal asymmetries (i.e., unsharped masking).  We excluded both the sub-clusters to avoid any bias on the model creation.  We show the masked regions in the top right panel of Figure \ref{fig:sw_circle}.
We found that a single beta model fits best for this Chandra image of the cluster. The best fit ellipticity of the cluster's counts image is found to be $0.120 \pm 0.002$ with its long axis (or semi-major axis) extending at $49.8 \pm 0.5$ degrees anti-clockwise with respect to the west direction. After determining the best-fitted parameters, we subtracted the model from the original image. The contours of the residual map extend $\sim 500$ kpc towards the South subcluster (top right panel of Figure \ref{fig:sw_circle}). As we masked two sub-cluster regions, the real extensions of the sloshing arm may vary in these regions. We estimated the effect of this sloshing arm by drawing an ellipse within the boundary of the sloshing gas. We found that the projected boundary of the sloshing arm coincides with the main filament region of the radio phoenix. We further discuss the implications of this in Sect. \ref{subsubsec:phoenix}.

\subsection{Shock Finder} \label{shock_finder}
We applied an automated 2D observational shock finder on the surface brightness and ACB temperature maps. This 2D shock-finder is inspired by a 3-D automated temperature-jump-based shock-finding algorithm developed by  \citet{Skillman2008,Skillman2011ApJ...735...96S,Skillman2013}. A detailed description of this 2D shock finder can be found in \citet{Datta2014,Hallman2018}.
It uses the Rankine–Hugoniot temperature jump condition (Equation \ref{eq:RH}) to calculate the Mach number.

\begin{equation}
\centering
    \frac{T_2}{T_1}=\frac{(5M^2-1)(M^2+3)}{16M^2}.
    \label{eq:RH}
\end{equation}

$T_1$ and $T_2$ are the pre-shock and post-shock temperatures, respectively. 
The shock Mach number is defined as $ M = \frac{v}{c_{s}} $, where, v is the velocity of shock and $ c_{s} = \sqrt{\frac{\gamma kT}{\mu m_{p}}} $ is the velocity of sound within the ambient ICM. 

This 2D shock-finding algorithm is inspired by the 3D shock-finding algorithm used in cosmological simulations by \citet{Skillman2008}. 
In the 3D shock finding algorithm, a pixel is determined to have a shock if it meets the following criteria. 
\begin{equation} \label{sv1}
    \div \Vec{v} < 0 
\end{equation}
\begin{equation} \label{sf2}
    \nabla T \cdot \nabla S > 0 
\end{equation}
\begin{equation} \label{st3}
    T_{2} > T_{1}
\end{equation}
\begin{equation} \label{sr4}
    \rho_{2} > \rho_{1}
\end{equation}
Where $\Vec{v}$ is the velocity field, T is the temperature, $\rho$ is the density, and S = T/$\rho^{\gamma-1}$ is the entropy. The 3D shock-finder calculates the jump in temperature and surface brightness at each pixel in both grid directions. A pixel is marked as a shock if the temperature gradient and surface brightness gradient have the same sign (as the temperature and density increase from pre-shock to post-shock). 
The Mach number is then calculated using Equation (1) from the temperature jump. 
Since the observations are restricted to the plane of the sky, \citet{Datta2014} had modified the 3D shock-finder to work on two-dimensional (2D) projected X-ray surface brightness and temperature maps. In this modified scenario, the shock-finder calculates the jump in temperature and surface brightness in N evenly placed directions centered on a given pixel. We used $N = 64$ for the shock map.
The shock-finder then accepts those pixel pairs between which the conditions $T_{2} > T_{1}$ and $S_{X2} > S_{X1}$ are satisfied (surface brightness is a proxy for density; in observations $\rho_2 > \rho_1$ was used to check the direction of the gradient). The other two conditions, $\div \Vec{v} < 0$ and $\nabla T \cdot \nabla S > 0$, cannot be used in the case of 2D real observations.
The Mach number for each successful pixel pair is noted. The Mach number with the maximum value is chosen to be the resultant Mach number for that given pixel. To choose the pixel pairs in each direction, we use a jump length from the ACB circle for that given pixel with a multiplicative factor. For A85, we also found that a multiplicative factor of 1.25 is appropriate for the Mach number estimation. 
Since our real observations are affected by projection in the 2D plane, applying equation \ref{eq:RH} to these observations will underestimate the Mach number. Due to this projection effect, the shock finder will produce a lower limit of the Mach numbers. However, the position of the shock will not change.

\begin{figure*}
\centering
\begin{tabular}{c|c}
    \includegraphics[width=0.95\columnwidth]{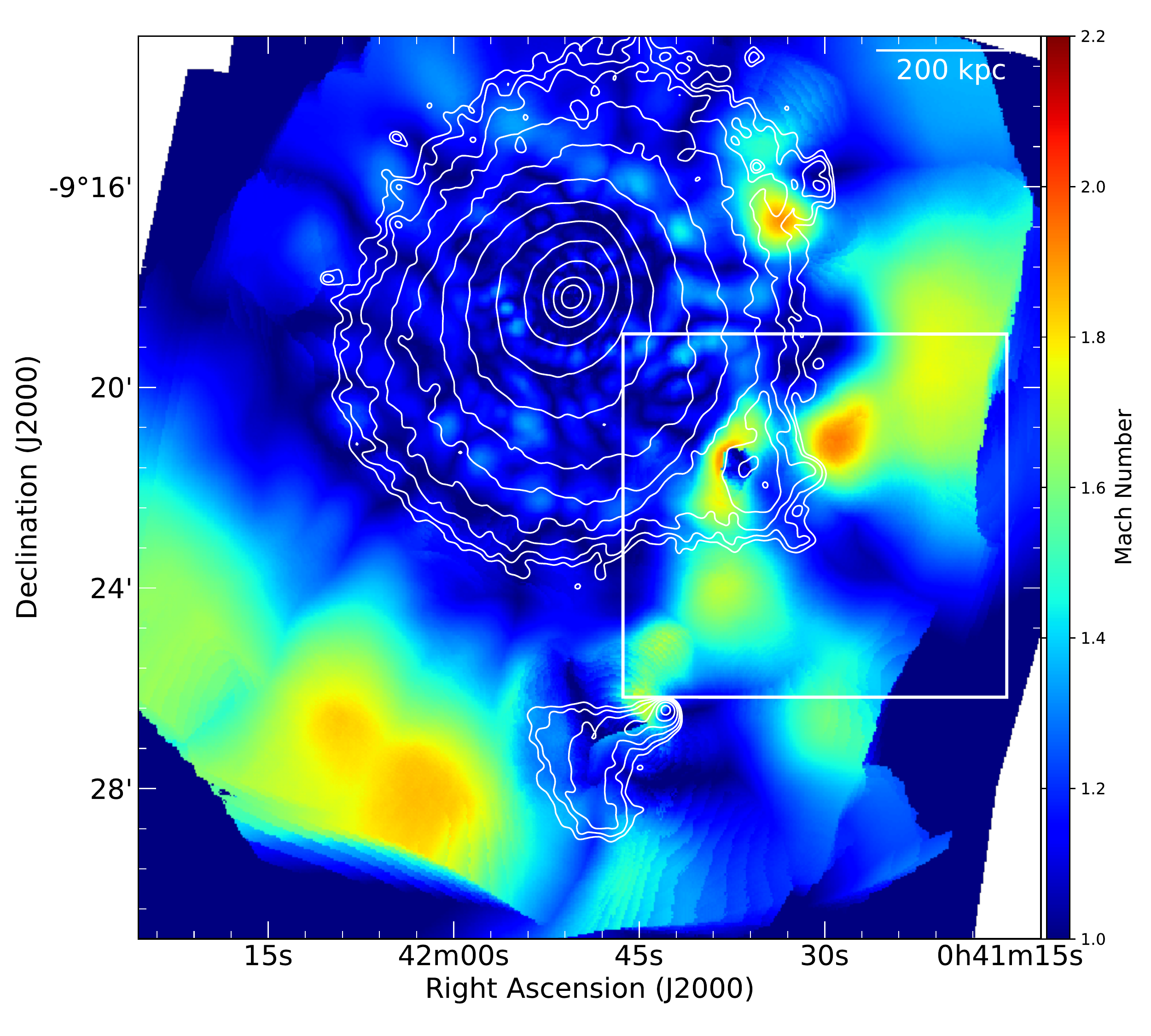} & \includegraphics[width=\columnwidth]{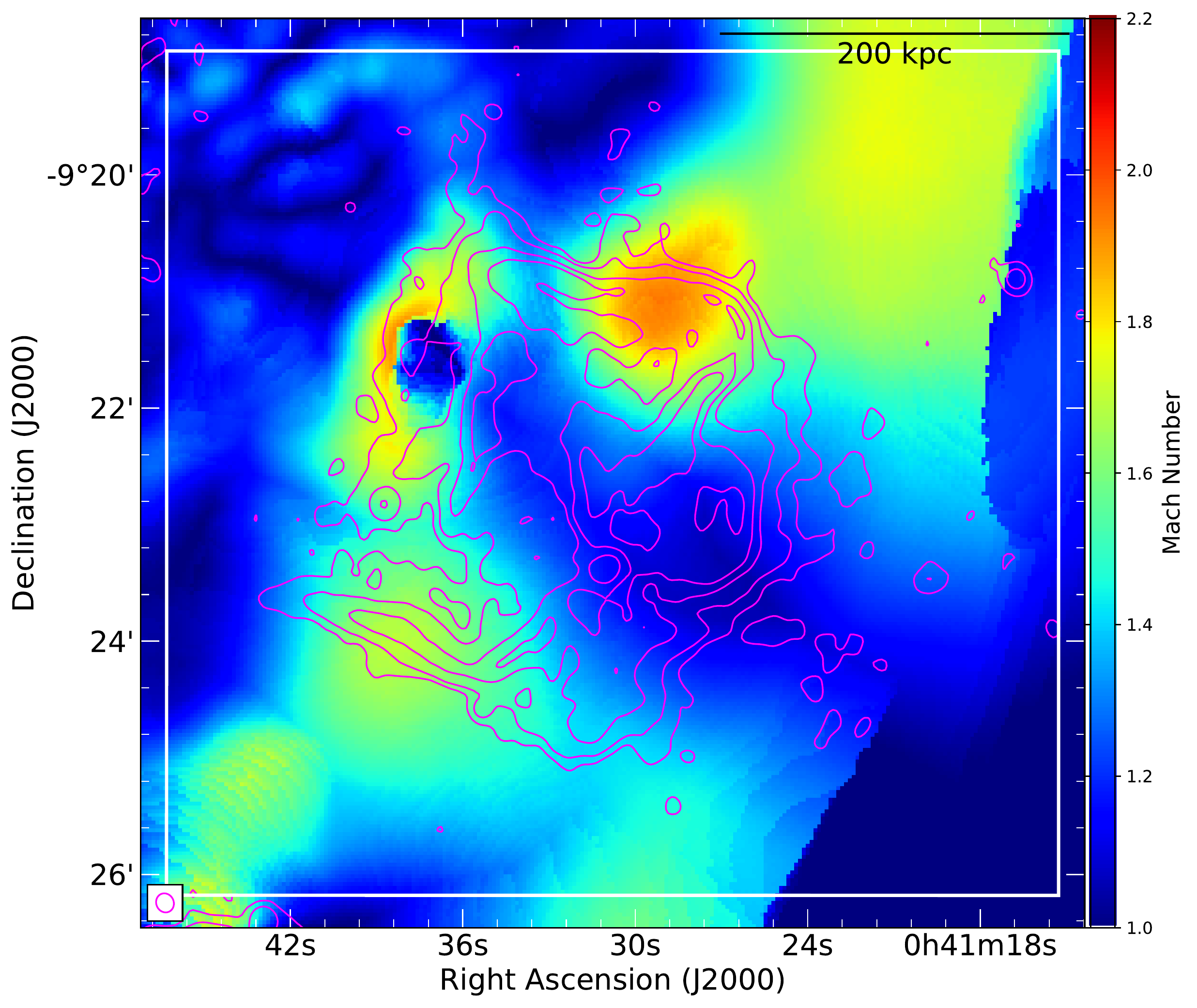}  \\
\end{tabular}
\caption{ Left: Mach number distribution map overlaid with the \textit{Chandra} X-ray surface brightness contours (white). The rectangular region contains the bow shock associated with the SW-subcluster. 
Right: A zoom-in of the box region from left panel is overlaid with the 325 MHz GMRT radio contour (magenta). The radio contours are drawn at levels $[1,2,4,8,16,32,64,128,256]\times 3\sigma_{rms}$, where $\sigma_{rms}$ = 96.2 $\mu$Jy beam$^{-1}$ and beam = $10 \arcsec \times 9.0 \arcsec$ and pointing angle 25.8 degree. }
\label{fig:4mach}
\end{figure*}

Using this shock finder, we produced a shock map (distribution of Mach numbers), which is shown in left panel of Figure \ref{fig:4mach}.
The automated shock finder identifies a bow shock in front of the southwest subcluster (white box region in Figure \ref{fig:4mach}).
In the bow shock region, the Mach number is found to be in the range of 1.4 to 1.9, which is consistent with the previous result from \citet{Ichinohe2015}. The position of the shock is also consistent with their study. 

\subsection{Improved radio images}
The full resolution 325 MHz GMRT radio observations produced the best-known radio image of A85 to date, where the clear filamentary structure of the radio phoenix can be seen. The filamentary structure is one of the main features distinguishing a radio phoenix from a radio relic. It should be noted that the current image has improved noise properties as well as resolution (10”x9”) compared to the previously published low frequencies image in \citet{Giovannini2000,Duchesne2017arXiv170703517D}. Moreover, this observation for the first time has revealed the extent of the diffuse emission in the radio phoenix up to 5 arcmin ($\approx 290$ kpc). 

\subsubsection{Radio Phoenix}
\begin{figure*}
\centering
\begin{tabular}{c|c}
    \includegraphics[width=3.1in]{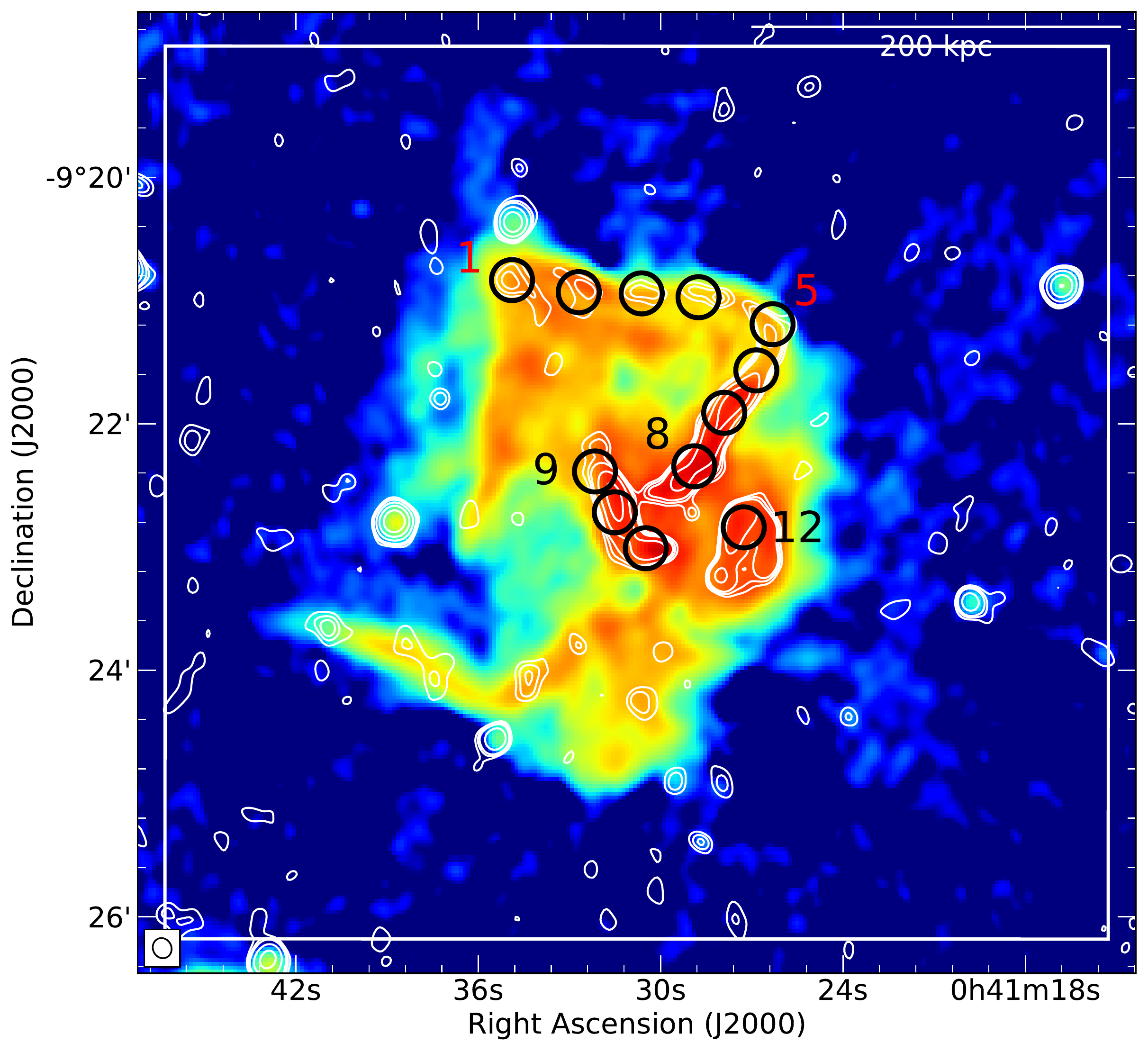} & 
    \includegraphics[width=\columnwidth]{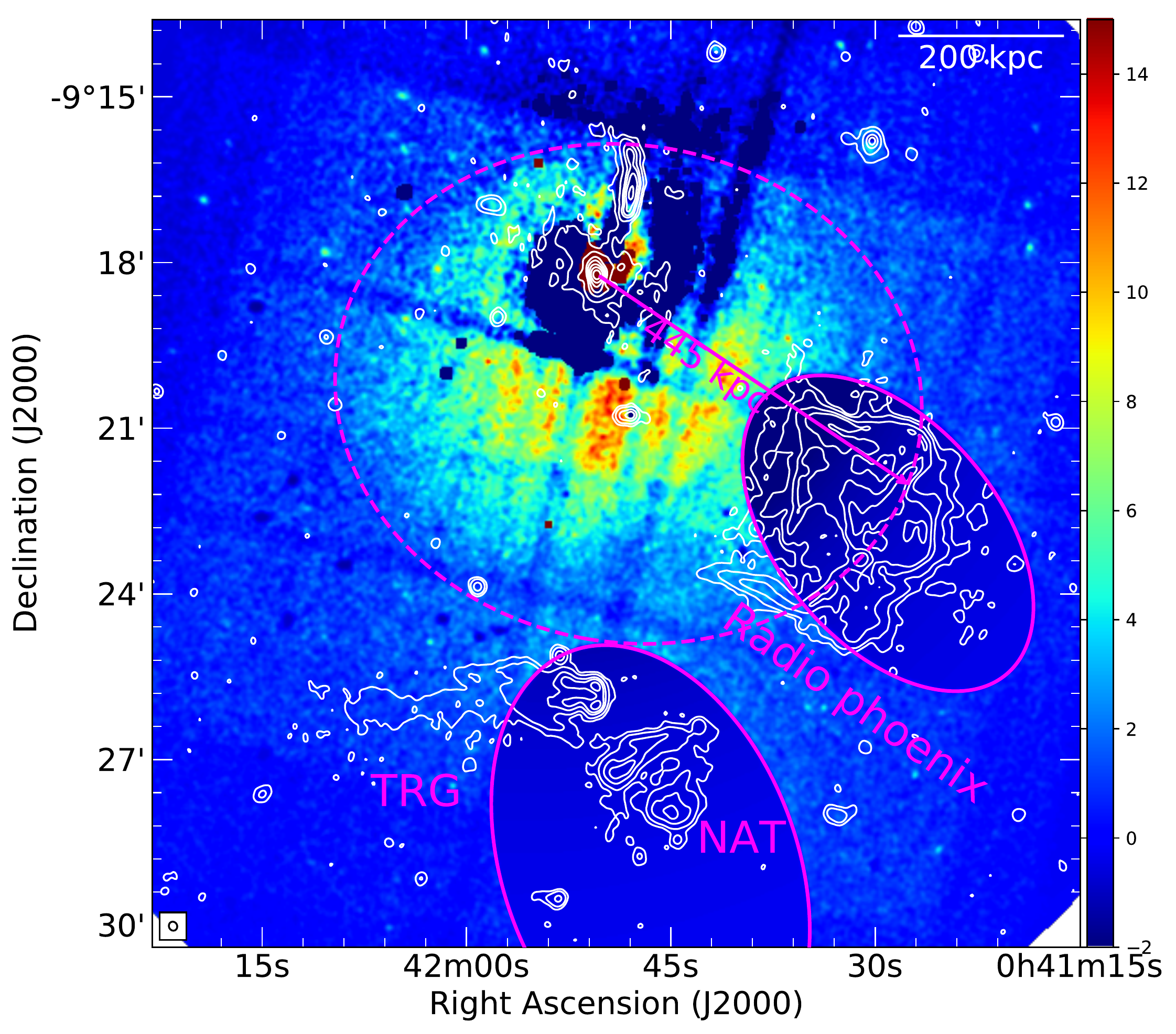} \\
    \includegraphics[width=\columnwidth]{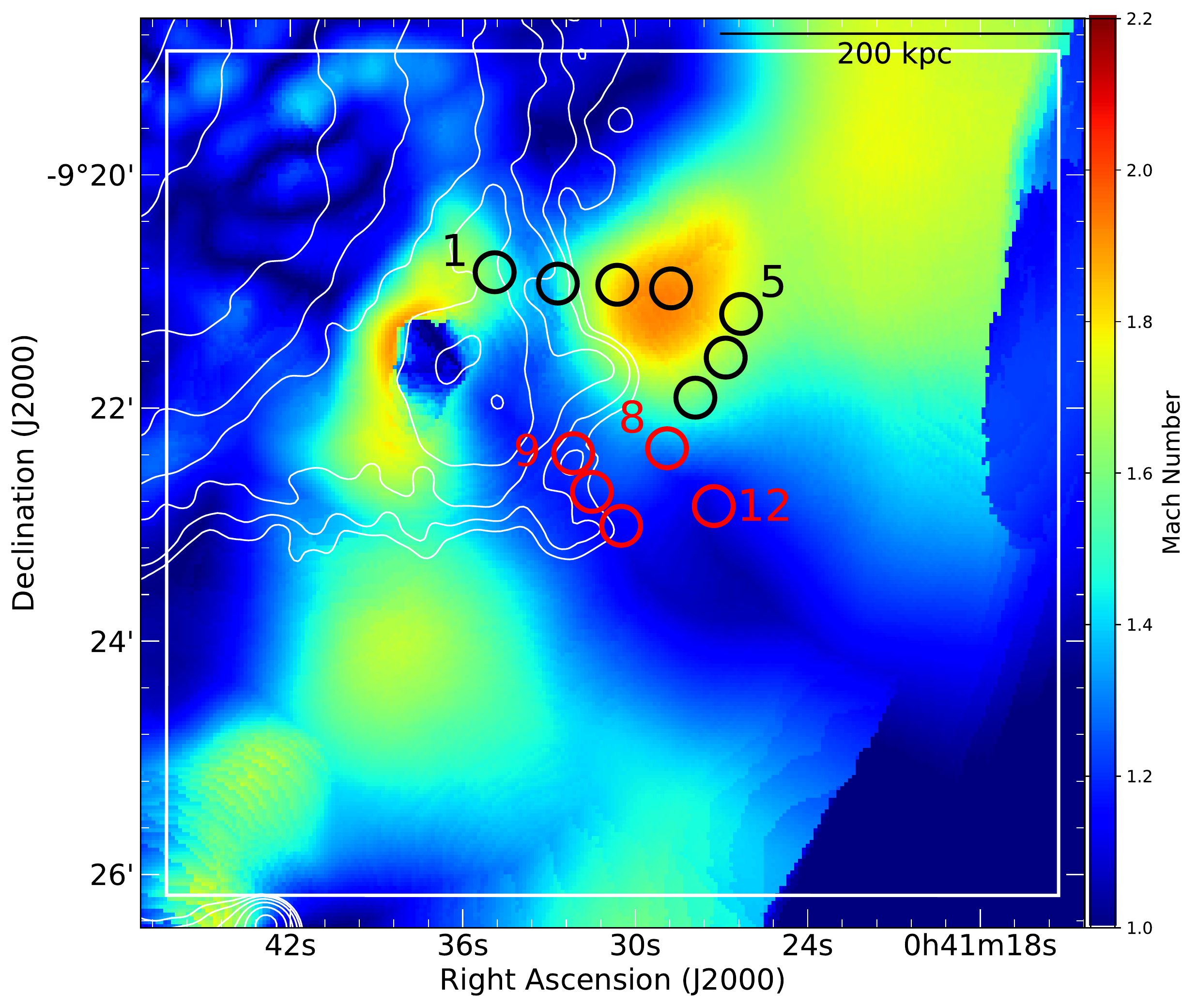} & \includegraphics[width=3.2in]{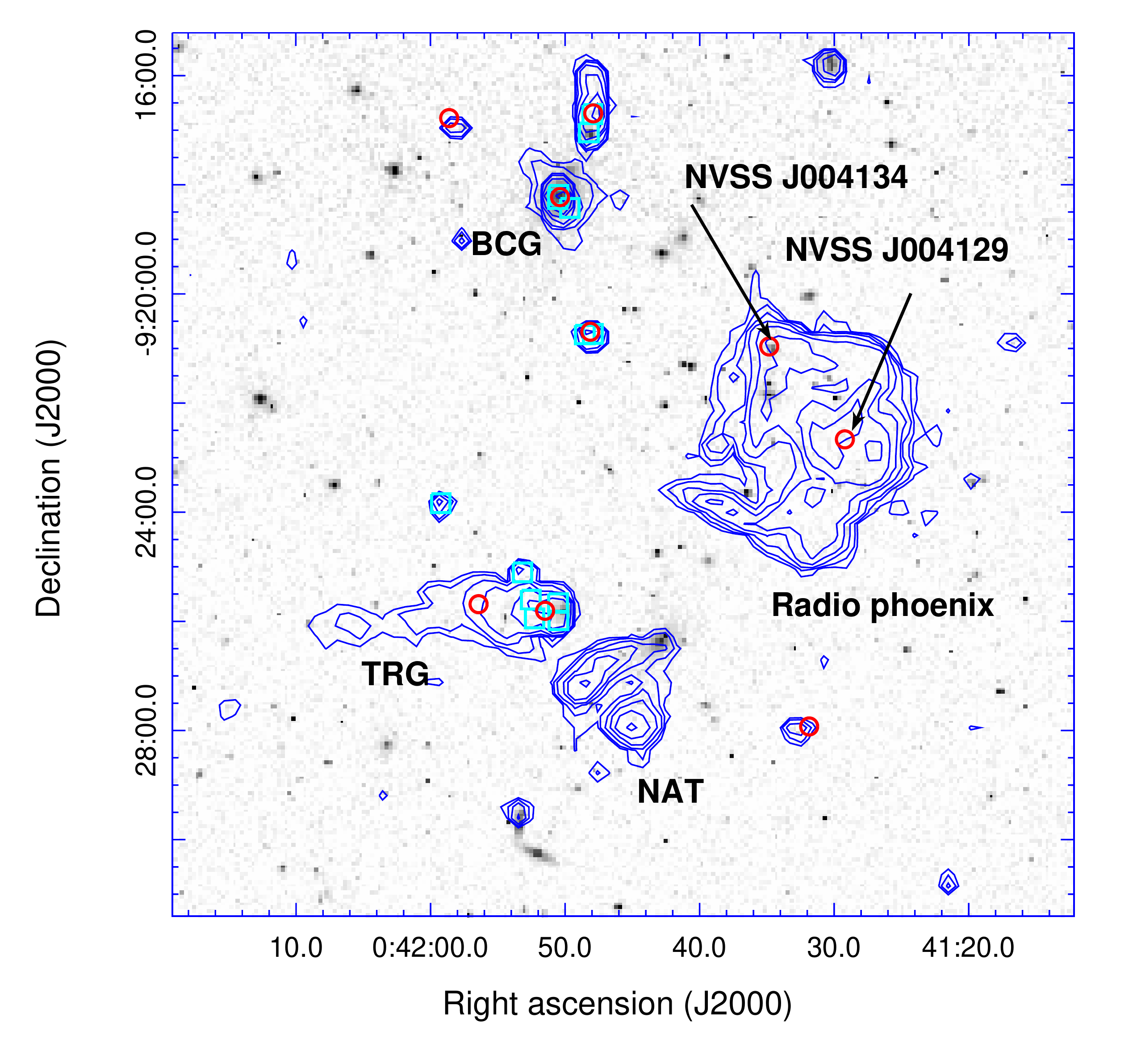} \\
    \end{tabular}
    \caption{Top Left: Radio phoenix shown in the 325 MHz GMRT radio band and overlaid with 1.4 GHz VLA L-band radio contours (white). The radio contours are drawn at levels $[-1,1,4,16,64,256]\times 3\sigma_{rms}$, where $\sigma_{rms}$ = 19.6 $\mu$Jy beam$^{-1}$ and beam = $10\arcsec \times 9\arcsec$. Both GMRT 325 MHz and VLA 1.4 GHz images have the same beam size as shown in the bottom left corner (cyan). Red and black circles with radius of $10 \arcsec$ (labeled as 1-12) are used to calculate the spectral index and Mach number (see Sect. \ref{radio_mach}). 
    Top Right: Chandra relative deviation image with respect to the best-fitting 2D elliptical double beta model overlaid with contours from the 325 MHz GMRT radio image (white). The radio contours are drawn at levels $[1,2,4,8,16,32,64,128,256]\times 3\sigma_{rms}$, where $\sigma_{rms}$ = 96.2 $\mu$Jy beam$^{-1}$ and beam = $10 \arcsec \times 9.0 \arcsec$ and pointing angle 25.8 degree.
    Two elliptical regions (magenta) shows the subcluster regions which were masked during the creation of 2D beta model. The rough estimated boundary of the sloshing arm is represented by the dotted ellipse (magenta). 
    Bottom Left: The zoomed-in Mach number distribution map (shock map; see Sect.  \ref{shock_finder}) was calculated using equation \ref{eq:RH} from ACB temperature maps, and is overlaid with \textit{Chandra} X-ray surface brightness contours (white).  
    Bottom Right: Optical image (SDSS r-band, DR12) of the A85 field overlaid with 325 MHz GMRT radio contours. The red circles represent sources from a catalog of the NVSS survey, and the cyan square box from a catalog of the FIRST radio survey. We found two radio-sources in the NVSS survey at the location of the radio phoenix within the cluster redshift range (e.g., NVSS J004134 or NVSS J004134-092057 and NVSS J004129 or NVSS J004129-092240).}
    \label{fig:sw_circle}
\end{figure*}

We present images of the radio phoenix in A85, using GMRT and VLA archival observations, in Figure \ref{fig:sw_circle}.
Filamentary structure of the radio phoenix is detected in the 325 MHz GMRT radio image (right panel of Figure \ref{fig:4mach}) at the location of the southwest subcluster.
The largest linear size (LLS) of this diffuse source is found to be $\thicksim$ 330 kpc at 325 MHz.
Only a small portion of the radio phoenix is visible in the higher frequency (1.4 GHz) image, where the 325 MHz emission is strong (see the white contours in Figure \ref{fig:sw_circle}).
It has an overall ultra-steep radio spectrum. 
The measured integrated flux densities of the radio phoenix are $S_{325\mathrm{MHz}} = 302.8 \pm 30.5$ mJy and $S_{1.4\mathrm{GHz}} = 10.3 \pm 1.1$ mJy (within the 3$\sigma$ contour).
The spectral indices listed in Table \ref{tab:spec_table} were calculated for some spatially independent circular regions (circles of 10$\arcsec$ radius, which are larger than the beam size; see Figure \ref{fig:sw_circle}) where radio emission from both frequencies is available.

\subsubsection{Radio Galaxies}
\begin{figure*}
\centering
\begin{tabular}{c|c}
\includegraphics[width=\columnwidth]{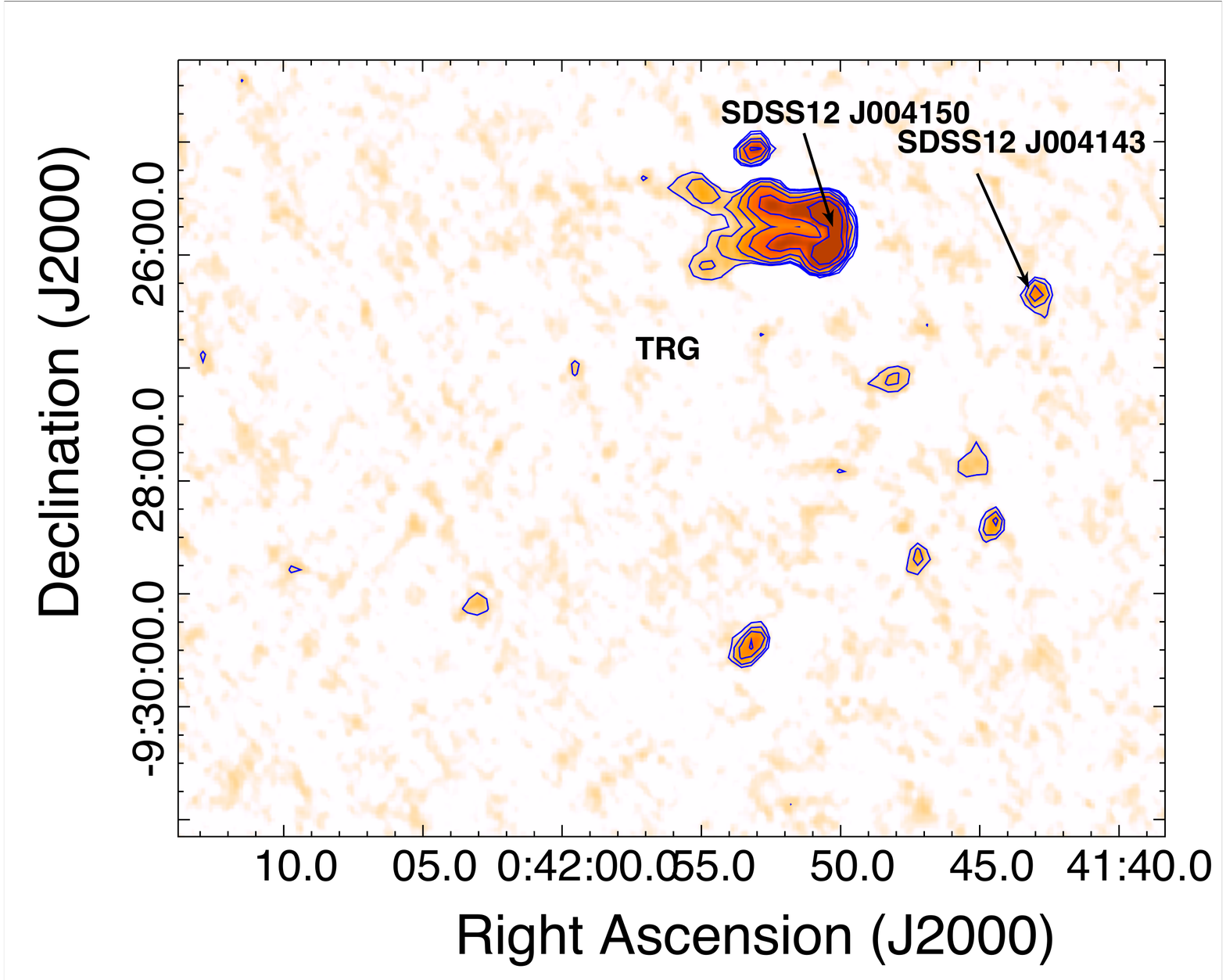} &
\includegraphics[width=\columnwidth]{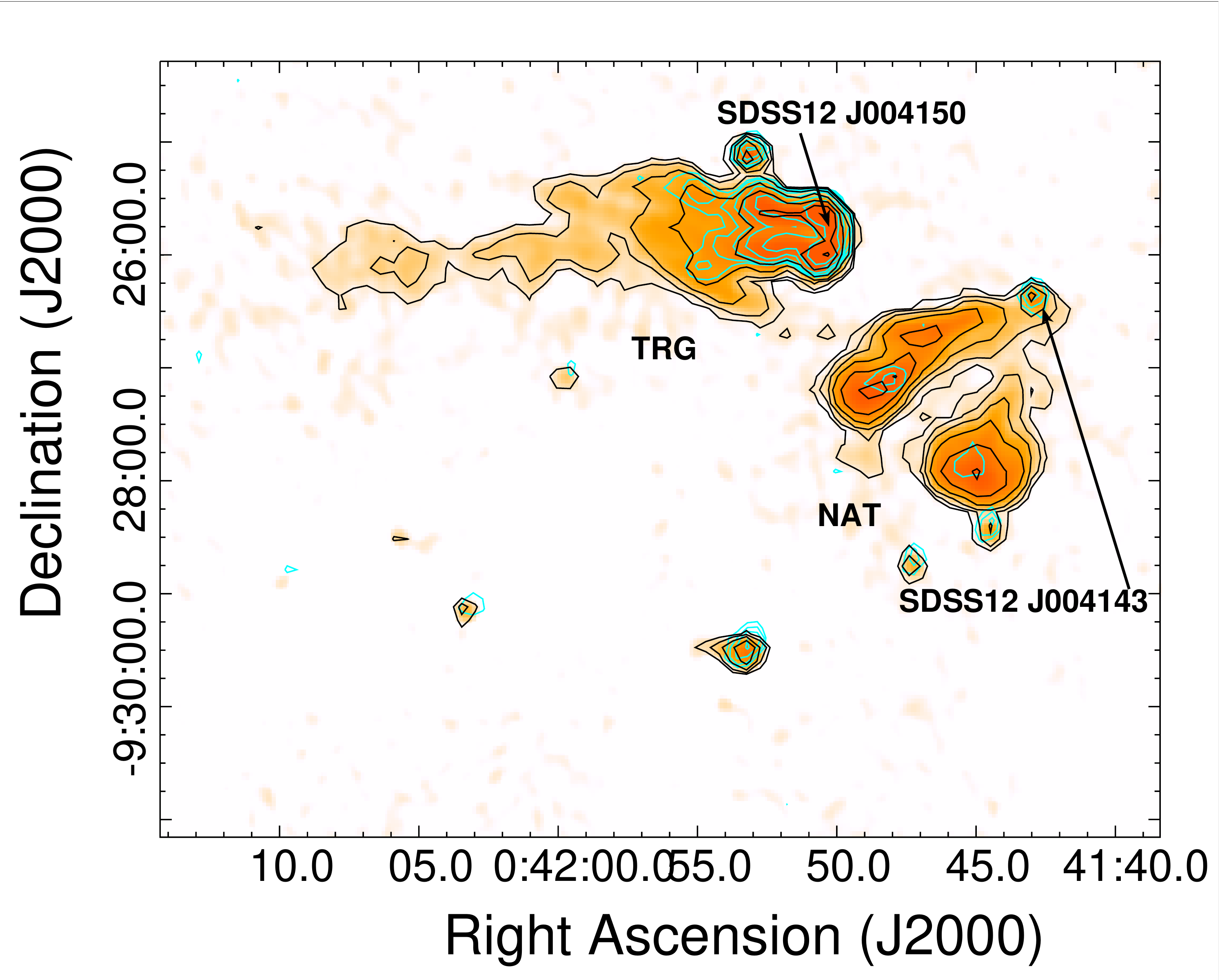} \\
\includegraphics[width=\columnwidth]{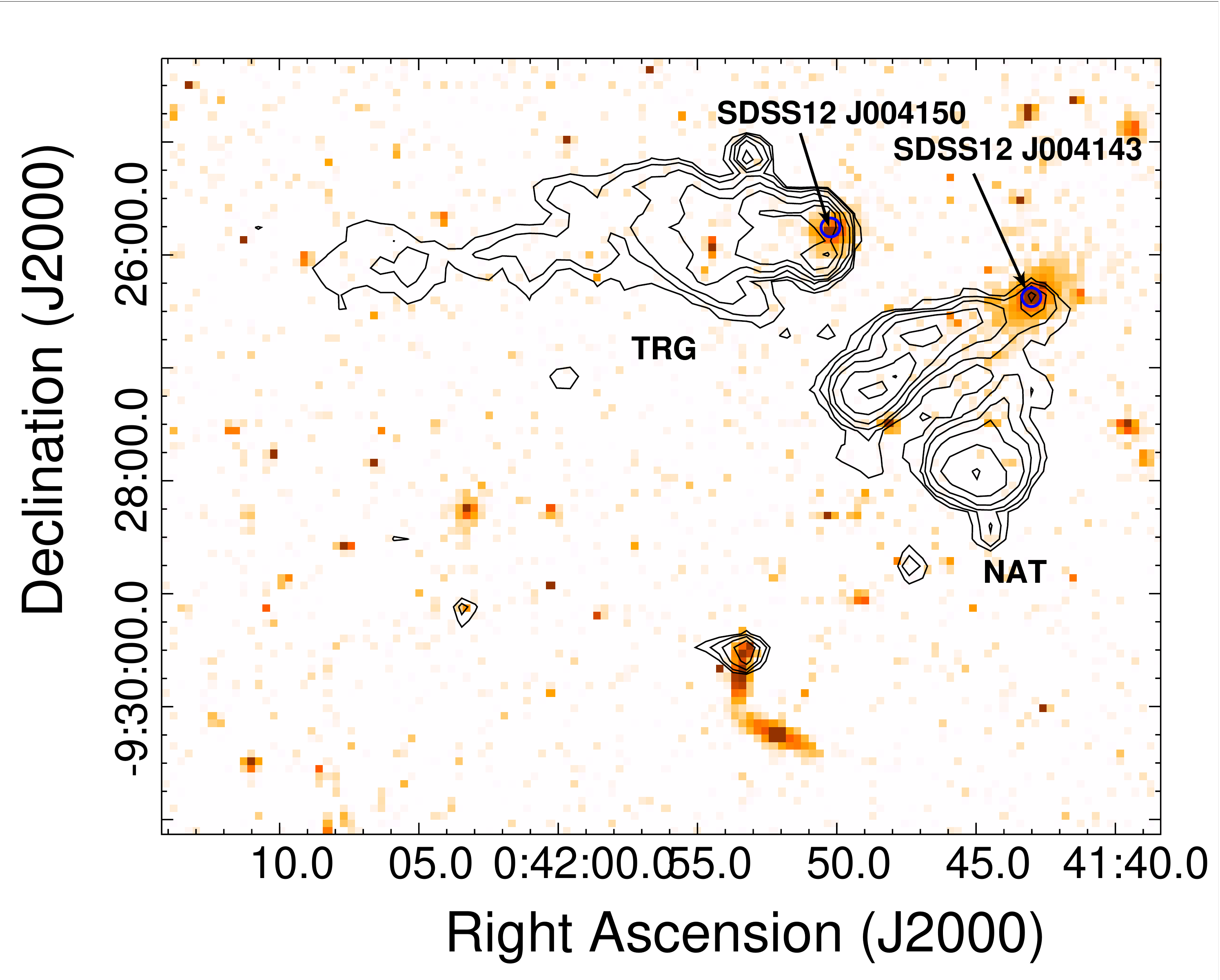} &
\includegraphics[width=\columnwidth]{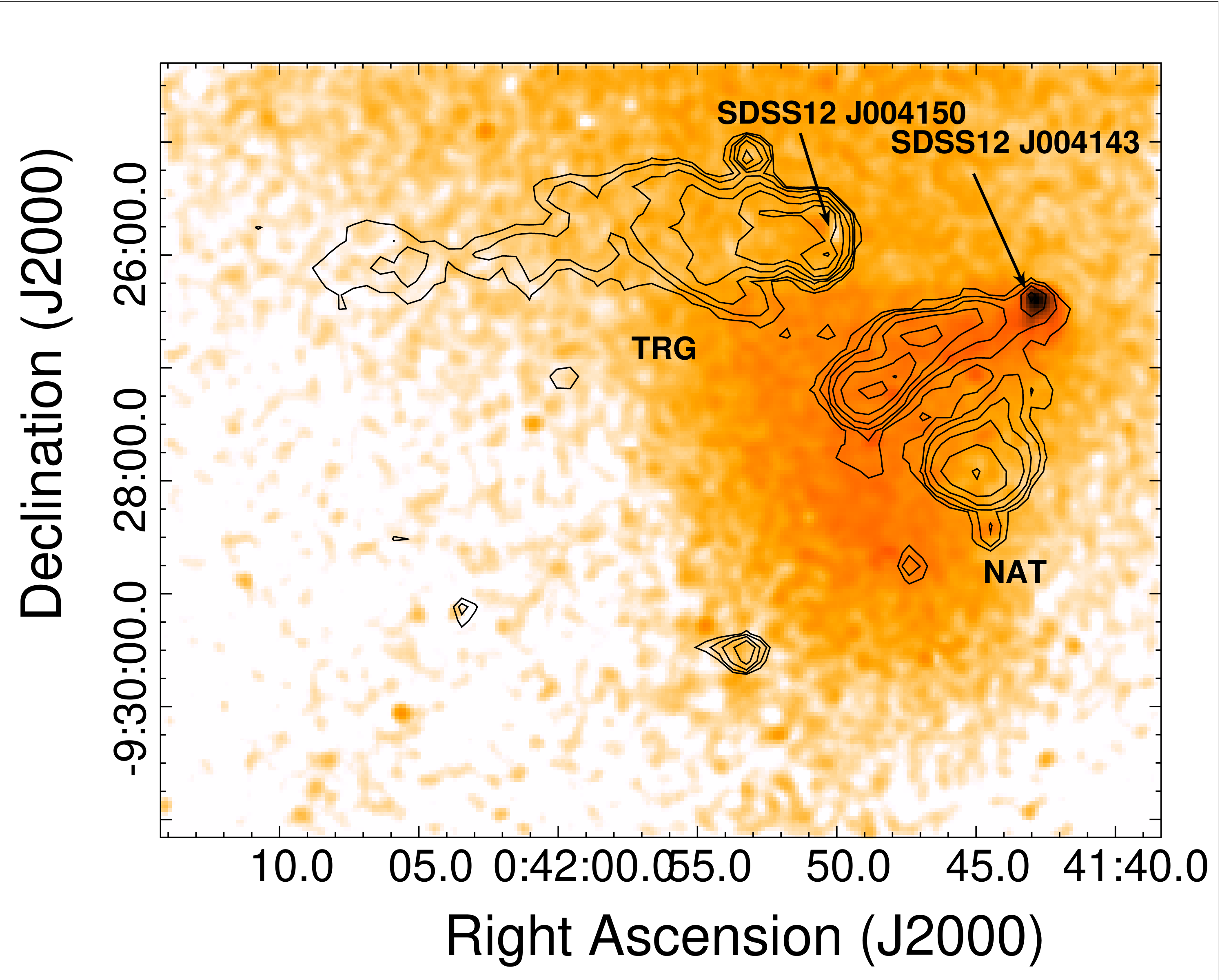} \\
\end{tabular}
\caption{
Top Left: 1.4 GHz VLA L-band radio image shown in colors and contours (blue). The radio contours are drawn at levels $[1,2,4,8,16,32,64,128] \times 3 \sigma_{rms}$, where $\sigma_{rms}$ = 19.6 $\mu$Jy beam$^{-1}$ and beam = $10\arcsec \times 9\arcsec$.
Top Right: 325 MHz GMRT radio image shown in colors and contours (black). The GMRT radio contours are drawn at levels $[1,2,4,8,16,32,64,128] \times 3\sigma_{rms}$, where $\sigma_{rms}$ = 96.2 $\mu$Jy beam$^{-1}$ and beam = $10\arcsec \times 9\arcsec$. This image was also overlaid with 1.4 GHz VLA L-band radio contours (cyan). The radio contours are same as shown in top left panel.
Bottom Left: The SDSS r-band optical image (DR12) of A85 overlaid with 325 MHz GMRT radio contours (black). The radio contours are same as top right panel. The blue circles are the position of the host galaxies from SDSS DR12.
Bottom Right: \textit{Chandra} X-ray surface brightness map of A85, overlaid with the 325 MHz GMRT radio contours (black). The contour levels are same as shown in the top right panel.
} 
\label{fig:galaxy}
\end{figure*}

A steep spectrum narrow-angle tail (NAT) radio galaxy is seen in the full resolution 325 MHz GMRT radio map (see Figure \ref{fig:galaxy}) at the position of the south subcluster. This radio galaxy was previously reported
by \citet{Giovannini2000} using 300 MHz VLA data. 
This narrow-angle tail (NAT) has an X-ray counterpart (bottom right, Figure \ref{fig:galaxy}) but was not visible at the L-band VLA image \citep{Slee2001,Schenck2014}.
The central part of this radio galaxy has a spectral index $\alpha_{RG}\thicksim -1.0$, the north tail has $\alpha_{NT}\thicksim -2.7$, and the south tail has $\alpha_{ST}\thicksim -2.8$. The rest of the radio emission is visible at 325 MHz only, and a rough upper limit to the spectral index is $\lesssim-3.3$. We estimated upper limits using the RMS noise from the same region of the L-band image.

Another tailed radio galaxy (TRG) is visible in both L-band VLA and 325 MHz GMRT images. Thanks to the high sensitivity of GMRT at 325 MHz, we have recovered the ultra-steep spectrum long tail of TRG for the first time. This TRG has no X-ray counterpart. 
The tail radio brightness is high in 325 MHz but not seen at 1.4 GHz. 
The total integrated flux density of this radio galaxy is found to be $142.0\pm14.0$ mJy in 325 MHz and $70.0\pm7.0$ mJy in 1.4 GHz.
The spectral index of the part of the radio galaxy visible at both frequencies is found to be $-0.5\pm0.2$. A rough spectral index upper limit of the tail was found to be $\lesssim-1.9$.

Both the radio galaxy have optical counterpart labeled as SDSS12 J004150.17-092547.4 (or SDSS12 J004150, z = $0.0566 \pm 0.0092$) for TRG and SDSS12 J004143.00-092621.9 (or SDSS12 J004143, z = $0.0560 \pm 0.0088$) for NAT. 

Detailed study of these two radio galaxies is out of the scope of this paper. Dedicated multi-frequency analysis on these two galaxies are in progress and will be published in another report.

\section{Discussions} \label{discussion}
\subsection{Bow shock}
A shock is created in a cluster environment when subclusters or groups of galaxies crush onto the cluster or pass through the cluster vicinity at supersonic velocities. In this process, gravitational potential energy converts into thermal energy, heating and compressing the ICM \citep{Roettiger1996,Burns1998,Datta2014}. There are two types of shocks: accretion shocks and merger shocks \citep{Datta2014}. Accretion shocks develop at large distances \citep{Skillman2008,Vazza2009} when cooler intergalactic gas accretes onto the cluster and produces high Mach number \textit{M} $\thicksim 10-100$ shocks \citep{Miniati2000,Ryu2003}.
A merger shock is produced when subclusters fall into the main cluster, resulting in lower Mach number (M $\leq$ 5) shocks \citep{Paul2011}.
The shock map of A85 (left panel of Figure \ref{fig:4mach}) reveals a bow shock propagating in front of the southwest subcluster at a distance of $\sim280$ kpc from the cluster core.
From the shock map, we found that the Mach number around the shock varies in the range of 1.4-1.9, which implies it is a weak merging shock.
This bow shock is exciting because it is situated in the vicinity of the radio phoenix present in the cluster.
We discuss the Mach number from both X-ray and multi-frequency radio observations below to investigate the relation between the X-ray shock and the radio phoenix.


\begin{table}
	\centering
	\caption{Radio spectral index ($\alpha$), corresponding Mach number calculated using Equation \ref{eq:mach_edge} and compared with the Mach number calculated using the Rankine–Hugoniot temperature jump condition (Eq. \ref{eq:RH}) from the ACB temperature map for the circular regions shown in Figure \ref{fig:sw_circle}.} 
	\label{tab:spec_table} 
	\begin{tabular}{lcccc} 
		\hline
        Reg & $M_X \pm $ Err & $\alpha$ $\pm$ Err & $M_{edge}$ $\pm$ Err & $M_{face}$ $\pm$ Err \\
        \hline
        1 & 1.5  $\pm$0.1 & -3.2  $\pm$0.5 & 1.3  $\pm$0.1 & 0.7  $\pm$0.3 \\
        2 & 1.3  $\pm$0.1 & -3.7  $\pm$0.9 & 1.3  $\pm$0.1 & 0.8  $\pm$0.2 \\
        3 & 1.1  $\pm$0.1 & -3.4  $\pm$1.0 & 1.3  $\pm$0.1 & 0.7  $\pm$0.2 \\
        4 & 1.9  $\pm$0.1 & -3.2  $\pm$0.6 & 1.3  $\pm$0.1 & 0.7  $\pm$0.3 \\
        5 & 1.7  $\pm$0.1 & -2.8  $\pm$0.4 & 1.4  $\pm$0.1 & 0.7  $\pm$0.3 \\
        6 & 1.6  $\pm$0.1 & -2.8  $\pm$0.3 & 1.4  $\pm$0.1 & 0.7  $\pm$0.3 \\
        7 & 1.5  $\pm$0.1 & -2.6  $\pm$0.1 & 1.42  $\pm$0.01 & 0.7  $\pm$0.3 \\
        8 & 1.3  $\pm$0.1 & -3.3  $\pm$0.2 & 1.31  $\pm$0.01 & 0.7  $\pm$0.3 \\
        9 & 1.23  $\pm$0.02 & -2.9  $\pm$0.1 & 1.42  $\pm$0.01 & 0.7  $\pm$0.3 \\
        10 & 1.21  $\pm$0.01 & -3.1  $\pm$0.3 & 1.33  $\pm$0.02 & 0.7  $\pm$0.3 \\
        11 & 1.12  $\pm$0.02 & -2.7  $\pm$0.1 & 1.41  $\pm$0.01 & 0.6  $\pm$0.3 \\
        12 & 1.11  $\pm$0.01 & -3.2  $\pm$0.2 & 1.32  $\pm$0.01 & 0.7  $\pm$0.3 \\
        \hline
	\end{tabular}
\end{table}

\subsection{Mach Number from Radio Observations} \label{radio_mach}
To estimate the spectral behavior of the diffuse radio emission in the southwest region, we calculated spectral indices between 325 MHz and 1.4 GHz for sample regions.
We chose these regions across twelve locations along the radio phoenix structure where radio emission is present in both radio frequencies (1.4 GHz and 325 MHz).
These consist of 12 circular regions (Figure \ref{fig:sw_circle}), each with a radius of $10\arcsec$, fairly sampling the radio structure covering both bright and less bright regions. 
The spectral index was calculated for each region, taking the sum of the flux encompassing the circles in both frequencies. 
The spectral index error was calculated using equation \ref{eq:spec_index_err} assuming the calibration error $\sigma_{\mathrm{cal}}$ in both frequencies to be 10\%,
\begin{equation}
    \Delta \alpha = \frac{1}{\log\Big(\frac{\nu_1}{\nu_2}\Big)}\sqrt{\Bigg(\frac{\Delta S_1}{S_1}\Bigg)^2 + \Bigg(\frac{\Delta S_2}{S_2}\Bigg)^2}
    \label{eq:spec_index_err}
\end{equation}
In A85, if we are viewing the diffuse emission “edge-on,” the radio spectral index ($\alpha$ defined as $S_{\nu} = \nu^{\alpha}$) should be sensitive to the prompt emission from the shock front \citep{Skillman2013} and is given by $\alpha = \alpha_{prompt} = (1 - s)/2$, where $s$ is the spectral index of the accelerated electrons are given by $n_{e}(E) \propto E^{-s}$ \citep{Hoeft_2007MNRAS.375...77H}. The theory of diffusive shock acceleration (DSA) for
planar shocks at the linear test-particle regime predict that
this radio spectral index is related to the shock Mach number by equation \ref{eq:mach_edge} \citep{Hoeft_2007MNRAS.375...77H,Ogrean2013}.
Alternatively, if the relic is viewed “face-on,” the radio spectral index is sensitive to cumulative spectral index ($\alpha = \alpha_{integrated} = - s/2$) due to emission from electrons over the entire lifetime. Hence, the corresponding Mach number is given by equation \ref{eq:mach_face} \citep{Skillman2013}.
\begin{equation}
\centering
    M^2_{edge}=\frac{2\alpha-3}{2\alpha+1}
	\label{eq:mach_edge}
\end{equation}
\begin{equation}
    M^2_{face}=\frac{\alpha + 1}{\alpha - 1}.
    \label{eq:mach_face}
\end{equation}
The spectral indices and corresponding Mach numbers are presented in Table \ref{tab:spec_table}. The Mach numbers calculated using the "edge-on" scenario are consistent with the Mach number ($M = 1.4 \pm 0.1$) reported in \citet{Ichinohe2015}.
This suggests a possible correlation between the diffuse radio emission and the X-ray shock. However, it should be noted that the location of the X-ray shock is not overlapping with the region where the radio Mach number is evaluated. This mismatch is due to the lack of surface brightness sensitivity in the VLA (1.4 GHz) image. Hence, the connection between the radio and X-ray Mach number is tentative and needs to be confirmed with high sensitivity 1.4 GHz observations.

\subsection{On the origin of Radio Phoenix} \label{subsubsec:phoenix}
The 325 MHz radio image shows a diffuse radio structure at the west of the southwest subcluster.
This diffuse radio structure was previously reported by  \citet{Joshi1986NRAOW..16...73J}, \citet{Bagchi1998} using 326.5 MHz OSRT radio data, by \citet{Giovannini2000} using 90 cm VLA radio data, and by \citet{Slee2001} and \citet{Schenck2014} using 1.4 GHz VLA radio data.
\citet{Kempner2004rcfg.proc..335K} and \citet{van_Weeren2019SSRv..215...16V} classified it as a radio phoenix. Radio phoenixes are, on average, smaller in size ($\leq 100-400$ kpc) and are found in smaller cluster centric distances \citep{van_Weeren2019SSRv..215...16V}. Radio phoenixes present a range of morphologies; from roundish to elongated and filamentary \citep{van_Weeren2019SSRv..215...16V}.
The full resolution radio map from the 325 MHz GMRT observation unraveled the complex structure of the radio phoenix in A85 (Figure \ref{fig:sw_circle}) at $\thicksim 280$ kpc from the center of the cluster. Our new high resolution 325 MHz radio image reveals the clear filamentary structure of the radio phoenix.
It consists of a both steep spectrum ($\alpha \thicksim$ -2.2) filamentary structure and a diffuse ultra-steep spectrum ($\alpha \thicksim$ -4.0) component.
The curved radio spectrum of this radio phoenix was previously reported by \citet{Slee2001} using a multi-frequency analysis.
The largest linear size (LLS) of this diffuse source is $\thicksim$ 330 kpc in the 325 MHz image.

The current favored scenario about the origin of the radio phoenix is that these radio sources trace old radio plasma from past AGN activity \citep{van_Weeren2019SSRv..215...16V}. When a shock passes through this old plasma, it compresses and increases the momentum of relativistic electrons and the magnetic field strength, producing a steep or curved radio spectrum source \citep{Enblin2001A&A...366...26E}.
A three-dimensional magneto-hydrodynamical simulation by \citet{Enblin2002MNRAS.331.1011E}, with the assumption of adiabatic compression of fossil radio plasma by a merger shock wave, predicted that these sources should have filamentary and complex morphology. 
 In one of their simulations, \citeauthor*{Enblin2002MNRAS.331.1011E} assumed a weak shock and weak magnetic fields, producing diffuse radio emission with a morphology (see Figures 1 and 6 in \citealt{Enblin2002MNRAS.331.1011E}) that matches that of the radio phoenix present in A85. 
Moreover, recent simulations \citep{Nolting2019ApJ...876..154N,Nolting2019ApJ...885...80N} showed that when cluster radio galaxy tails pass through ICM shocks or large relative motions, they induce filamentary morphologies and ultra-steep spectra, similar to what we observe in A85.

\citet{Mandal2019arXiv191102034M} reported radio phoenixes in three clusters (Abell 2593, Abell 2048, and SDSS-C4-DR3-3088), and proposed the scenario that phoenixes might be connected with shocks and ICM motions.
We re-investigated the shock present at the front of the SW subcluster by creating a shock map, which confirms the presence of an X-ray shock (see Sect. \ref{shock_finder}), which was 
first reported by \citet{Ichinohe2015}. From the shock map (bottom left panel of Figure \ref{fig:sw_circle}), we also confirm the X-ray bow shock position, which is situated towards the east edge of the radio phoenix.
The Mach number was estimated from multi-frequency radio maps and compared with the Mach number
from the shock map (see Table \ref{tab:spec_table}) as well as from the Mach number reported in \citet{Ichinohe2015}. 
The Mach number from the radio (\lq\lq edge-on\rq\rq\ Mach number) and X-ray data are consistent with each other (see Sect. \ref{radio_mach}), which implies that the radio phoenix may have resulted from compression of the fossil plasma bubble by the X-ray shock.
It should be noted that there are some caveats about comparing radio and X-ray Mach numbers. In this case, the radio emission from L-band VLA observations covers only a small portion of the radio phoenix, slightly away from the X-ray shock. X-ray Mach numbers are calculated from the projected X-ray temperature maps, which can also underestimate the Mach numbers. The projection effect can affect both the radio and X-ray observations. 

Large scale gas sloshing motion can affect the origin of radio emission in clusters.
Recent magneto-hydrodynamic simulations of cluster mergers by \citet{ZuHone_2021ApJ...914...73Z} show that merger-driven gas motions (a) can advect bubbles of cosmic rays to very large radii, and (b) can spread relativistic seed electrons producing extended, filamentary, or sheet-like regions of intracluster plasma enriched with aged cosmic rays, which resemble radio relics.  Since radio phoenixes are small-scale radio relics that are present near to the cluster center in contrast to the conventional large-scale relics, we investigated this gas sloshing scenario via the residual map (see Figure 3 and Sect. \ref{residual}). 
We found a large spiral gas sloshing arm extending up to $\sim500$ kpc from the cluster centre. 
The extension of the sloshing arm differs slightly from the previous study by \citet{Ichinohe2015}. This might be because they did not mask the subcluster region and therefore, their residual map was effected by contributions from the sub-clusters.
The "arm" starts from the north of the core and extends counter-clockwise outward from the core.
The morphology of the sloshing arm indicates that the interaction plane is close to the plane of the sky. The morphology of the radio phoenix also suggests that the fossil plasma might be concentrated mostly in the southwest direction.

From Figure \ref{fig:sw_circle}, the filamentary region of the radio phoenix appears coincident with the boundary of the sloshing arm. This coincidence implies that the flow of the ICM resulting from sloshing might spread the cloud of the fossil plasma towards the southwest direction. As the radio phoenix present in A85 is situated $\sim$300 kpc away from the center, it is not favorable for the central BCG to be the source of bubble relativistic electrons. We searched for an off-axis radio galaxy as the origin of a bubble of pre-existing relativistic electrons and found only two radio-sources (galaxies) within the radio phoenix region (see the bottom right panel of Figure \ref{fig:sw_circle}).
Therefore, it is possible that ram pressure from the sloshing gas may be responsible for the internal plasma distribution producing the radio phoenix structure shown in the top right panel of Figure \ref{fig:sw_circle}. 
When the SW-subcluster passes through the bubble of relativistic electrons, it may light up the plasma bubble, producing large scale filamentary radio emission - the radio phoenix in A85.

\section{CONCLUSION} \label{sec:conclude}
In this paper, we analyzed radio observations in two different bands, 325 MHz from GMRT and 1.4 GHz from VLA. We also re-analyzed archival Chandra X-ray observations of A85 and made an ACB-smoothed temperature map and an X-ray shock map.
Results from this work are summarised as follows:

\begin{itemize}
\item We presented new 325 MHz GMRT radio observations (see Figure \ref{fig:sw_circle}) that show detailed filamentary structure of the radio phoenix present in the cluster. This is the highest quality image of the radio phoenix in A85 to date. The filamentary structure is one of the main features of a diffuse radio emission seen in a radio phoenix. It should be noted that our current image has significantly improved noise properties as well as resolution (10”$\times$9”) compared to  previously published low frequency images in \citet{Giovannini2000,Duchesne2017arXiv170703517D}.

\item  We investigated the possible origin of the radio phoenix in A85 in Sect. \ref{subsubsec:phoenix}. 
We speculate that the most favorable scenario for the formation of the radio phoenix might be a merging shock with a low mach number ($<$ 2) that can compress ghost plasma bubbles and re-energize the electrons to boost their visibility at frequencies below few hundred MHz 
(e.g., \citealt{Enblin2001A&A...366...26E,Enblin2002MNRAS.331.1011E,Nolting2019ApJ...876..154N,Nolting2019ApJ...885...80N}).
Radio emission resulting from this compression has a filamentary morphology and ultra-steep or curved spectrum. Another possibility is that the bubbles of relativistic electrons injected by jets from the off-center radio galaxy NVSS J004134-092057 were spread by large-scale gas sloshing present in A85 and lit up radio emission when the SW subcluster or any shock related to the SW-subcluster crosses this region.

\item The radio observations also divulged an ultra-steep spectrum narrow-angle tail galaxy (NAT, in Figure \ref{fig:sw_circle}) within the south subcluster that is only visible in the low frequency 325 MHz images.
Another bright-tailed radio galaxy (TRG) at the northeast of the south subcluster was also detected (see TRG in Figure \ref{fig:sw_circle}) in both radio images. The long tail of this tailed radio galaxy is only visible at 325 MHz and has an ultra-steep spectrum.

\item The \textit{Chandra} ACB temperature map of A85 reveals a complex, disturbed morphology. 
Such substructure in temperature maps are signatures of merging activity. 
The overall temperature structure is consistent with previous studies by \citet{Schenck2014,Ichinohe2015}.

\end{itemize}

\section*{Acknowledgements}
MR and RR would like to thank Manoneeta Chakraborty, Arnab Chakraborty, Bhargav Vaidhya for fruitful discussions.
MR acknowledges the support provided by the DST-Inspire fellowship program (IF160343) by DST-Inspire, India. This work is supported through ECR/2017/001296 grant awarded to AD by DST-SERB, India. This work was also supported by NASA ADAP grant NNX15AE17G to JB. Through DR, this work was also supported by NASA (National Aeronautics and Space Administration) under award number NNA16BD14C for NASA Academic Mission Services.
We thank IIT Indore for providing computing facilities for data analysis. We also thank both GMRT and VLA staff for making these radio observations possible. GMRT is run by the National Centre for Radio Astrophysics of the Tata Institute of Fundamental Research. The scientific results from X-ray reported in this article are based on data obtained from the \textit{Chandra}. This research has also made use of the software provided by the Chandra X-ray Center (CXC) in the application packages CIAO, ChIPS, and Sherpa. This
research made use of APLpy \citep{aplpy2012,aplpy2019}, an open-source plotting package for Python.

\section*{DATA AVAILABILITY}
The X-ray data underlying this article are available in \textit{Chandra} data archive at [https://cda.harvard.edu/chaser/mainEntry.do].
The radio data used in this article are available in GMRT archive at [https://naps.ncra.tifr.res.in/goa/data/search].
The derived data presented in this article will be shared on reasonable request to the corresponding author.




\bibliographystyle{mnras}
\bibliography{reference,references2} 




\appendix
\section{Weighted Voronoi Tessellation}   \label{wvt}
\begin{figure}
\centering
\includegraphics[width=\columnwidth]{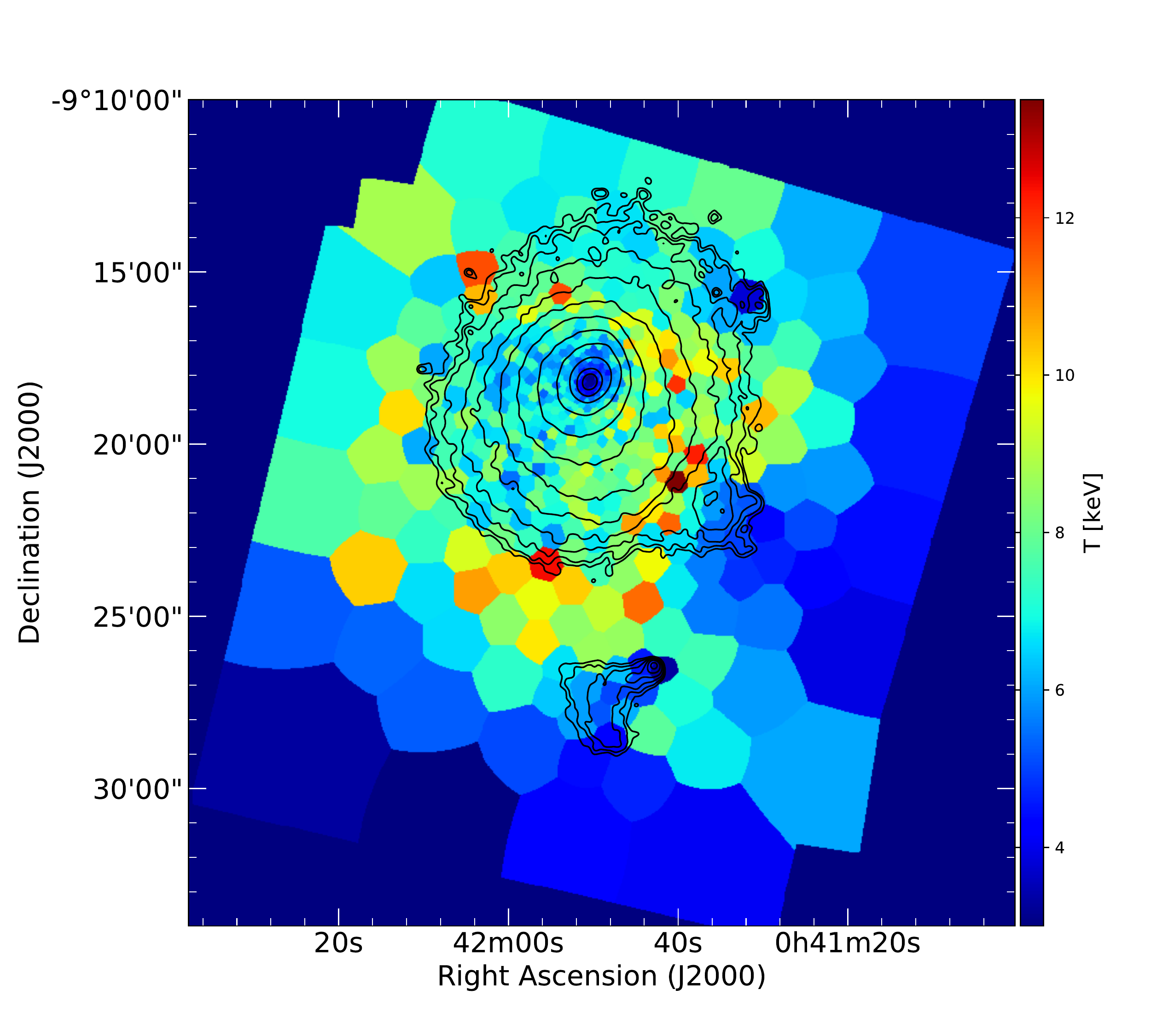} 
\caption{\textit{Chandra} X-ray WVT temperature map of A85 in the 0.7-8 keV energy band overlaid with X-ray surface brightness contours. Here, we used a signal to noise ratio (SNR) of 50, which gives 539 WVT regions.}
\label{fig:WVT}
\end{figure}

This method was developed by \citet{Diehl2006} and widely used to derive temperature maps (e.g., \citealt{Pratt2007,Simionescu2007,Randall2008,Gastaldello2009,Rossetti2010}) of clusters of galaxies (e.g., A4059; \citealt{Reynolds2008,Mernier2015}, A3128; \citealt{Werner2007d}, A2052; \citealt{Blanton2009,Plaa2010}, A2744 \citealt{Owers2011}, A2254; \citealt{Girardi2011}, A1201; \citealt{Ma2012}, 1RXS J0603.3+4214; \citealt{Ogrean2013}, CenA; \citealt{Walker2013d}, A85; \citealt{Schenck2014}, A3667; \citealt{Datta2014}, NGC 1404; \citealt{Su2017}, Fornax; \citealt{Su2017b}).
The advantage of the WVT method is that it produces statistically independent and spatially non-overlapping regions. The ACB temperature map has a higher resolution than that from WVT but with overlapping regions.
SNR is calculated using merged CLEAN data and CLEAN background. Here, we used a SNR of 50, which comprises 539 WVT regions. 
After creating these regions, we used the \textit{dmextract} \textit{CIAO} task to extract spectra from both source and background for each observation separately for the same region. The weighted response matrix (RMF) and weighted effective area (ARF) are calculated using \textit{specextract} for each observation. 
We performed spectral fitting using XSPEC version: 12.9.1 within the 0.7-8.0 keV energy range. The APEC and PHABS models were fitted to the spectra of each region. We fitted all five spectra for each region from 5 ObsId's simultaneously. C-Statistics \citep{Cash1979} was used for spectral fitting. Metalicity (0.3 $Z_\odot$), redshift (0.0556) and $N_H$ of the cluster were kept frozen.  Only APEC normalization and temperature were fitted for each region. The temperature map is obtained and presented in Figure \ref{fig:WVT}.

\bsp	
\label{lastpage}
\end{document}